\documentclass[journal=jacsat,manuscript=article]{achemso}
\usepackage[version=3]{mhchem} 

\usepackage{graphicx}
\usepackage{dcolumn}
\usepackage{amssymb}
\usepackage{color}
\usepackage{epstopdf}
\usepackage{float}

\usepackage{tikz,xcolor,hyperref}
\definecolor{lime}{HTML}{A6CE39}
\DeclareRobustCommand{\orcidicon}{
	\begin{tikzpicture}
	\draw[lime, fill=lime] (0,0) 
	circle [radius=0.16] 
	node[white] {{\fontfamily{qag}\selectfont \tiny ID}};
	\draw[white, fill=white] (-0.0625,0.095) 
	circle [radius=0.007];
	\end{tikzpicture}
	\hspace{-2mm}
}
\foreach \x in {A, ..., Z}{
    \expandafter\xdef\csname orcid\x\endcsname{\noexpand\href{https://orcid.org/\csname orcidauthor\x\endcsname}{\noexpand\orcidicon}}
}

\author{Cindy Liza Esporlas\orcidA{}}
\author{Georgiy Tkachenko\orcidB{}}
\author{Isha Sanskriti\orcidC{}}
\author{Viet Giang Truong\orcidD{}}
\author{S\'{i}le {Nic Chormaic}\orcidE{}}
\email{sile.nicchormaic@oist.jp}
\affiliation{Okinawa Institute of Science and Technology Graduate University, Onna, Okinawa 904-0495, Japan}

\title{Efficient Optical Manipulation of Janus Particles by Optical Nanofibers}


\begin{document}

\begin{abstract}
  Small composite objects, known as Janus particles, drive sustained scientific interest primarily targeted at biomedical applications, where such objects act as micro- or nanoscale actuators, carriers, or imaging agents. The major practical challenge is to develop effective methods for manipulation of Janus particles. The available long-range methods mostly rely on chemical reactions or thermal gradients, therefore having mediocre precision and strong dependency on the content and properties of the carrier fluid. To tackle these limitations, we propose the manipulation of Janus particles (here, silica microspheres half-coated with gold) by optical forces in the evanescent field of an optical nanofiber. We find that Janus particles exhibit stronger transverse localization and faster propulsion compared to all-dielectric particles of the same size. The propulsion speed recorded for a 3-$\mu$m particle with a 20-nm-thick gold cap averages at 2~$\mu$m/s per 1~mW of optical power, reaching 133 body length/s at 200~mW going through the nanofiber.
\end{abstract}


\noindent The term Janus particle (JP), inspired by the ancient Roman two-faced god, was coined in 1988 by C.~Casagrande {\it et al.} to describe microscopic beads with differently treated hemispheres, one made polar (hydrophilic), and the other apolar (hydrophobic)~\cite{casagrande_epl_1989}. The term was then adopted by the scientific community in a broader sense, as a descriptor for a composite micro- or nanoscale artificial object having multiple parts with distinct chemical or physical properties. The great potential of such particles was acknowledged in 1991 by P.-G.~de~Gennes in his Nobel Prize lecture on Soft Matter~\cite{deGennes_s_1992}. Indeed, JPs sparked intense research activity which led to numerous applications, primarily in the biomedical domain because of the versatility and size-wise compatibility of such particles with living tissue~\cite{yi_analyst_2016, le_intJNanoMed_2019}. Some notable examples include the use of JPs for unidirectional association with human endothelial cells~\cite{yoshida_advMat_2009}, magnetolytic therapy for cancer~\cite{hu_jacs_2010}, specific cellular targeting, sensing, and spectroscopy~\cite{wu_small_2010}, and biomarking for detection via dark-field imaging~\cite{sotiriou_chemMat_2011}. In recent years, the development of JPs shows a trend toward eco-friendly and sustainable materials~\cite{marschelke_colPolSci_2020}. Although the most common geometry for a JP is a sphere with distinct halves, many other kinds are possible, e.~g. dimers~\cite{valadares_small_2010}, nanocorals~\cite{wu_small_2010}, nanotrees~\cite{dai_natNano_2016}, and even homogeneous particles having transient ``Janus'' properties induced by an external field~\cite{chen_advMat_2017, schmidt_nc_2021}. Typically, a JP is produced via nanofabrication, by incorporating a nanoscale metallic coating or inclusion in a homogeneous carrier particle, usually a commercially available dielectric bead. 

Naturally, one would wish to have a precise tool for handling JPs, similar to what optical tweezers are for contactless manipulation of various small-scale objects~\cite{gao_lsa_2017}. After first tests in optical tweezers~\cite{merkt_njp_2006}, it was soon noted that JPs with metallic parts are very hard to handle by light, because the high reflectance and absorbance of the metal lead to strong optical and thermal forces repelling the particle from the optical trap. For instance, JPs in water were seen to escape higher-intensity regions and settle around higher-gradient ones~\cite{jiang_prl_2010, mousavi_softMat_2019}. It was also recognized that -- precisely because of their inhomogeneity -- JPs commonly exhibit various forms of ``taxis'', that is, motion in response to external stimuli, such as  temperature~\cite{jiang_prl_2010, bickel_pre_2013, bickel_pre_2014, lozano_nc_2016, wu_nanoRes_2016, mousavi_softMat_2019}, diffusion~\cite{volpe_softMat_2011, ghosh_prl_2013},  magnetic fields~\cite{dong_acsNano_2016}, or chemical reactions~\cite{valadares_small_2010, ma_nanoLett_2015, dai_natNano_2016, dong_acsNano_2016, wang_langmuir_2017, chen_advMat_2017, singh_advMat_2017}. Phototaxis via thermophoretic forces emerged as one of the most promising methods, because it allows one to move the particles by light, without the need for any chemical fuel. However, these forces  are very weak in common environments ({\it e.~g.}, water) and fail to produce considerable speed. To enhance thermophoresis and thus make a JP an efficient microswimmer or micromachine, major attention has been given to studies on metallo-dielctric JPs in near-critical mixtures of liquids where the system's behavior is at its most sensitive to minute changes in temperature or pressure~\cite{volpe_softMat_2011, buttinoni_jcmp_2012, wurger_prl_2015, lozano_nc_2016, schmidt_nc_2021}.

Unrestricted JPs assume random orientations in the host environment. Therefore, special arrangements have to be made in order to make the induced motion directional. In particular, one can use confined environments~\cite{volpe_softMat_2011, ghosh_prl_2013}, ``polarization" in thermal gradients~\cite{bickel_pre_2014}, gravity~\cite{nedev_acsPhot_2015}, structural defects~\cite{zong_acsNano_2015,wang_langmuir_2017}, magnetic fields~\cite{dong_acsNano_2016}, elastic forces~\cite{simoncelli_acsPhot_2017}, multiple programmable light sources~\cite{dai_natNano_2016, chen_advMat_2017} or structured fields~\cite{lozano_nc_2016}. With optical techniques being the most promising, until now there have been no efficient methods for fast and precise manipulation of JPs via optical forces alone. Such a method would be very much beneficial for many applications of JPs, because a purely optical manipulation does not rely on chemical processes or special conditions such as temperature distributions, exotic mixtures, or fluidic backgrounds.

Recalling that metallo-dielectric JPs in optical fields tend to move toward higher-gradient areas~\cite{jiang_prl_2010, mousavi_softMat_2019}, a logical strategy is to use evanescent fields, where the intensity exhibits steep gradients near the surface, thereby constraining the particle in the radial direction while allowing it to move along the wavevector. There have been numerous works on optical manipulation via evanescent fields in the vicinity of a glass prism~\cite{kawata_ol_1992}, planar waveguide~\cite{kawata_ol_1996}, and more recently, ultrathin optical fibers  mediating propulsion~\cite{brambilla_ol_2007, lei_LabOnChip_2011, frawley_oe_2014,li_OptFibTech_2016,yoshino_oe_2020}, binding~\cite{frawley_oe_2014, gusachenko_phot_2015, maimaiti_sr_2016, toftul_acsPhot_2020}, rotation~\cite{tkachenko_optica_2020}, detection~\cite{sugawara_oe_2020}, and sorting~\cite{fujiwara_SciAdv_2021, sadgrove_arxiv_2021} of various kinds of micro- and nanoparticles.

In this study, we report on the first theoretical and experimental analysis of all-optical manipulation of JPs in the evanescent field of an optical nanofiber~(ONF). Our particles are silica microspheres, half-coated with a few-nanometer-thick layer of gold. We compare the longitudinal motion and transverse localization of coated and uncoated particles, and discuss such factors as the gold cap's orientation and thermal effects.

\section{Results and discussion}
\subsection{Optomechanical model}
\begin{figure}
\centering\includegraphics[width=\linewidth]{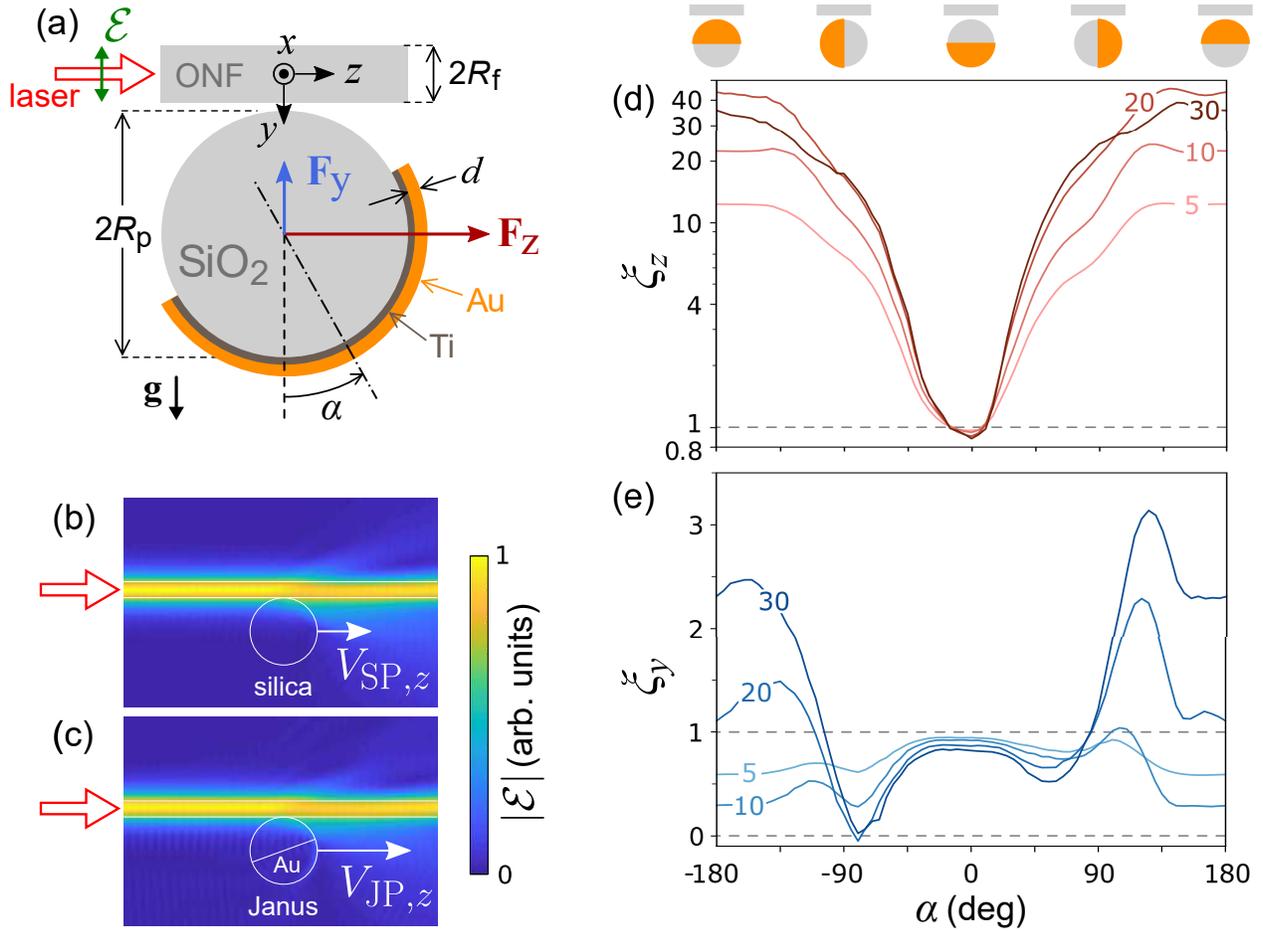}
\caption{(a)~Model schematics for a gold-silica Janus particle in the evanescent field near an optical nanofiber guiding a single mode polarized in the $yz$ plane. (b,~c)~Simulated electric field distributions around silica and Janus ($d=20$~nm, $\alpha=20^{\circ}$) particles. (d,~e)~Simulated enhancement factors for (d)~longitudinal and (e)~radial optical forces acting on a gold-coated silica particle, in relation to those for an uncoated silica particle of the same size. The numbers on the curves indicate the gold cap thickness, $d$, in nanometers. The titanium coating is 5-nm-thick for all particles. Top: gold cap orientations at every 90~degrees.}
\label{fig:figComsol}
\end{figure}

We consider optical forces acting on a silica-gold JP in water near an ONF guiding a single plane-polarized mode. Since both the spherical JP and the cylindrical ONF+mode are mirror-symmetric objects, we expect the combined light-matter system in its stationary state to be symmetric with respect to the polarization plane, given the gravitational acceleration, $\bf g$, also lies in this plane. Therefore, we can use the simplified two-dimensional model sketched in Figure~\ref{fig:figComsol}a. Here, the mode propagates along the fiber axis, $z$, the input laser beam has the electric field vector, ${\bf \cal E}$, lying in the $yz$ plane, and $\bf g$ is parallel to~$y$.

The system has the following parameters: the vacuum wavelength of the laser is $1,064\,\mu$m; the ONF and the particle have diameters of $2R_{\rm f}=0.7\,\mu$m and $2R_{\rm p}=3\,\mu$m, respectively, and are surrounded by water at normal ambient conditions; one hemisphere of the particle is coated with a thin uniform layer of gold over a 5-nm-thick titanium adhesion layer~\cite{rakic_ao_1998}. The particle is assumed to be in contact with the fiber (up to a 10-nm-thick gap due to the surface roughness~\cite{marchington_oe_2008}) under the action of the radial optical force, ${\bf F}_y$, and moving along the fiber by the longitudinal optical force, ${\bf F}_z$. In order to straightforwardly connect the forces and the resultant dynamics of the particles, we define dimensionless {\it enhancement} factors,
\begin{equation}
    \xi_{i=y,z} = F_{{\rm JP},i} / F_{{\rm SP},i}\approx V_{{\rm JP},i} / V_{{\rm SP},i}\,,
    \label{eq:xi}
\end{equation}
where $F_{{\rm JP},i}$ and $F_{{\rm SP},i}$ are respectively the magnitudes of the optical forces on Janus and silica~(SP) particles, and $V_{{\rm JP},i}$ and $V_{{\rm SP},i}$ are the corresponding speeds of the particle's centre-of-mass. The approximate equality in Eq.~\ref{eq:xi} relies on the assumption that coated and uncoated particles have nearly identical geometries, and that the speed is proportional to the driving force (in the limit of small Reynolds number as is the present case). The latter statement implies that thermal changes to the fluid's viscosity are negligible.

The two variable parameters for the gold cap are its thickness, $d$, and the orientation angle, $\alpha$, defined as the tilt of the particle's symmetry line with respect to the axis~$y$. As shown in Figure~\ref{fig:figComsol}d,e, these parameters significantly affect the enhancement factors for both longitudinal (Figure~\ref{fig:figComsol}d) and radial (Figure~\ref{fig:figComsol}e) directions. Noteworthy, $\xi_i(\alpha)$ curves are not symmetric with respect to $\alpha=0^{\circ}$, as one might expect. This is due to the guided mode breaking the $z\leftrightarrow (-z)$ reflection symmetry of the ONF+JP system.

For each given thickness, $d$, the propulsion enhancement, $\xi_z$, is maximal when the gold cap is oriented toward the fiber ($|\alpha|\to 180^{\circ}$). This result is quite intuitive, given the relatively high polarizability of gold and the sharp decline in the evanescent field intensity as a function of distance from the fiber surface. In the explored range of gold thickness from 5 to 60~nm, the maximum $\xi_z$ approaches~50 and saturates around $d = 20$~nm, which is close to the skin depth of gold at the working wavelength of light~\cite{svoboda_ol_1994}. When the cap is facing away from the fiber ($|\alpha|\approx0^{\circ}$), the enhancement is close to unity, meaning that the gold layer with this orientation has only a minor effect on the light-induced propulsion of the particle.

As will be shown in the experimental section, our JPs are usually being propelled with $\alpha \sim 20^{\circ}$, which corresponds to a modest propulsion enhancement of $1<\xi_z<4$. The faster propulsion of a JP, $V_{{\rm JP},z}>V_{{\rm SP},z}$, is related to the slightly stronger back-scattering it produces ({\it cf.} bottom-left parts of Figure~\ref{fig:figComsol}b and Figure~\ref{fig:figComsol}c).

For the radial direction, the maximum enhancement is much smaller than for the longitudinal one. Some orientations of the cap give $\xi_y\ll1$ or even $\xi_y<0$ (for larger $d$ values and $\alpha \approx-80^{\circ}$), in which case the radial optical force on the particle is repelling it from the fiber. Unfortunately, these simulations do not allow us to predict the orientation angle, $\alpha$, and the corresponding light-induced dynamics of JPs in the evanescent field. The only way to accurately assess these dynamics is through experiment and, for this purpose, we chose two representative values of gold thickness of 20~nm and 10~nm. The former is expected to produce the strongest propulsion enhancement, while the latter provides a quantitative verification of the model by allowing us to compare two distinct cases. The choice of $d=5$~nm would provide even higher distinction; however, in practice such thin gold coatings appeared patchy following JP fabrication and therefore were not suitable for making comparisons with the model geometry.

\begin{figure}
\centering\includegraphics[width=\linewidth]{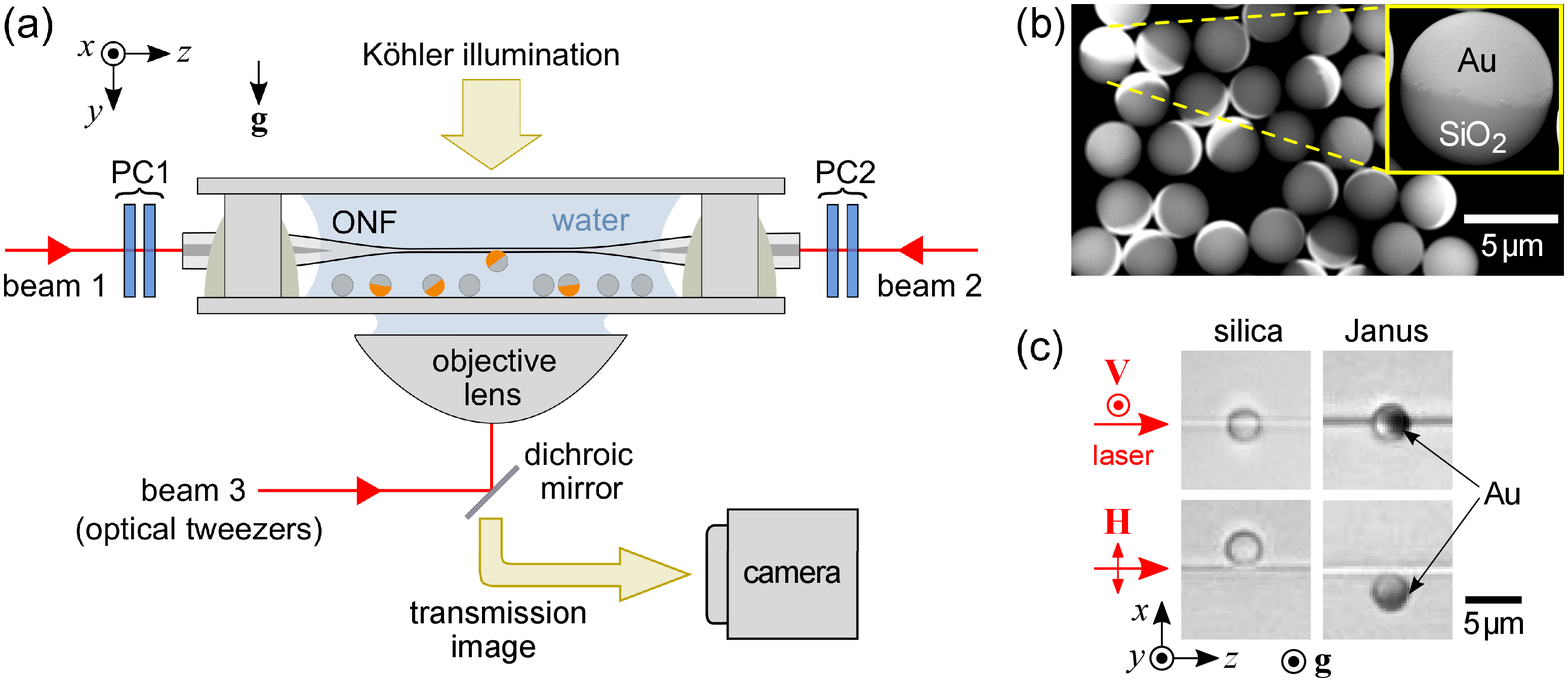}
\caption{Experimental details. (a)~Schematic of the optical setup (not to scale). PC1,2: polarization compensators, consisting of two quarter-wave plates each. (b)~Typical SEM images of detached Janus particles made by half-coating 3-$\mu$m silica spheres with a 20-nm-thick layer of gold. (c)~Camera images of a Janus particle (right) and a silica particle (left)  optically trapped in the evanescent field of an ONF, for 100~mW of transmitted optical power and either $\bf V$ (top) or $\bf H$ (bottom) polarization in beam~1.}
\label{fig:figSetup}
\end{figure}

\subsection{Experiment}
Since the optomechanical behavior of the Janus particles was considered in relation to that of silica particles, we performed each set of measurements using a mixture of both coated and uncoated particles interacting with the same short stretch of the waist region of an optical nanofiber (Figure~\ref{fig:figSetup}a). We originally planned to disperse particles in heavy water (D$_2$O) since it has a negligible absorption at the working wavelength of 1.064~$\mu$m. However, this choice proved impractical, because JPs in D$_2$O kept escaping from our single-beam optical tweezers. In contrast, in Milli-Q water, we found that a JP was quickly attracted to the beam focus and hovered around it under the combined action of optical, gravitational, and thermophoretic forces, of which the last evidently required the fluid (not only the particle) to have some absorption. Even in this case, the JPs were only stably trapped and delivered to the ONF when the optical power in the tweezers did not exceed 3~mW (Supporting Information Video~S1); otherwise the particle  wandered further from the focus and eventually escaped due to Brownian motion. It was particularly challenging to manipulate JPs in the $xz$ plane, because of the weakness of the  optical gradient forces. Aside from the slightly darker region of the gold cap (Figure~\ref{fig:figSetup}c), we used this unusual behavior as a clear indication that a given particle was a JP rather than a SP, which predictably exhibited stable optical trapping with the trap stiffness being proportional to the optical power.

\begin{figure}
\centering\includegraphics[width=\linewidth]{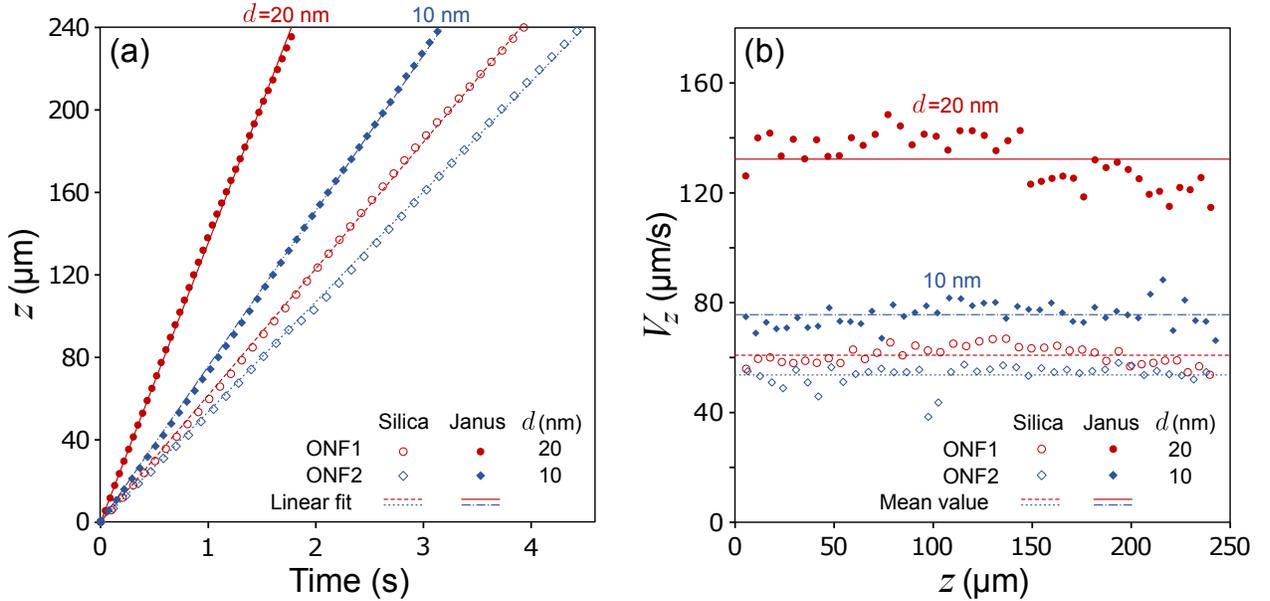}
\caption{Measured optomechanics of $3~\mu$m silica and Janus particles near an ONF guiding a single mode polarized in the $yz$ plane ($\bf V$ input polarization) for a transmitted optical power of~80~mW. Two samples were used: ONF1 for $d=20$~nm and and ONF2 for $d=10$~nm.  (Supporting Information Video~S2). (a)~Position-time series for the longitudinal direction of the particle motion. (b)~Propulsion speed vs. particle position along the ONF waist. Mean speed values in $\mu$m/s: $\langle V_{{\rm JP},z} \rangle|_{\rm ONF1} = 132.4$, $\langle V_{{\rm SP},z} \rangle|_{\rm ONF1} = 61.0$, $\langle V_{{\rm JP},z} \rangle|_{\rm ONF2} = 75.7$, and $\langle V_{{\rm SP},z} \rangle|_{\rm ONF2} = 53.8$.}
\label{fig:figTraj}
\end{figure}

Figure~\ref{fig:figTraj}a shows a typical longitudinal position-time series for a JP and a SP recorded at a transmitted power through the ONF of 80~mW (see Methods for more details on the power estimation) and $\bf V$ polarization in  beam~1 (hence, $yz$ polarization at the ONF waist). Markers indicate the $z$ coordinate of the particle's center (with respect to that in the first frame), with an approximate spacing of 6~$\mu$m. As follows from the increased steepness of the position-time series, the JP move significantly faster than the SP in the corresponding sample, labelled as ONF1 for a gold thickness of 20~nm and ONF2 for 10~nm. The slight divergence between the series for SPs in the two ONF samples suggests that our methods of ONF fabrication and finding the optimum region of its waist for the measurements are not exactly reproducible. However, the well-controlled measurement procedure and the relative analysis of the motion (via $\xi_i$ rather than $V_i$) enable us to minimize the systematic errors in the experiments. Apart from the observed propulsion speed enhancement ($\xi_z>1$), we find that both uncoated and gold-coated particles exhibit a \emph{linear} position-time series (Figure~\ref{fig:figTraj}a), and nearly \emph{constant longitudinal speeds} (Figure~\ref{fig:figTraj}b). Therefore, our model assumption that thermal effects due to the gold cap are negligible is largely supported by the experimental evidence. 

\begin{figure}
\centering\includegraphics[width=\linewidth]{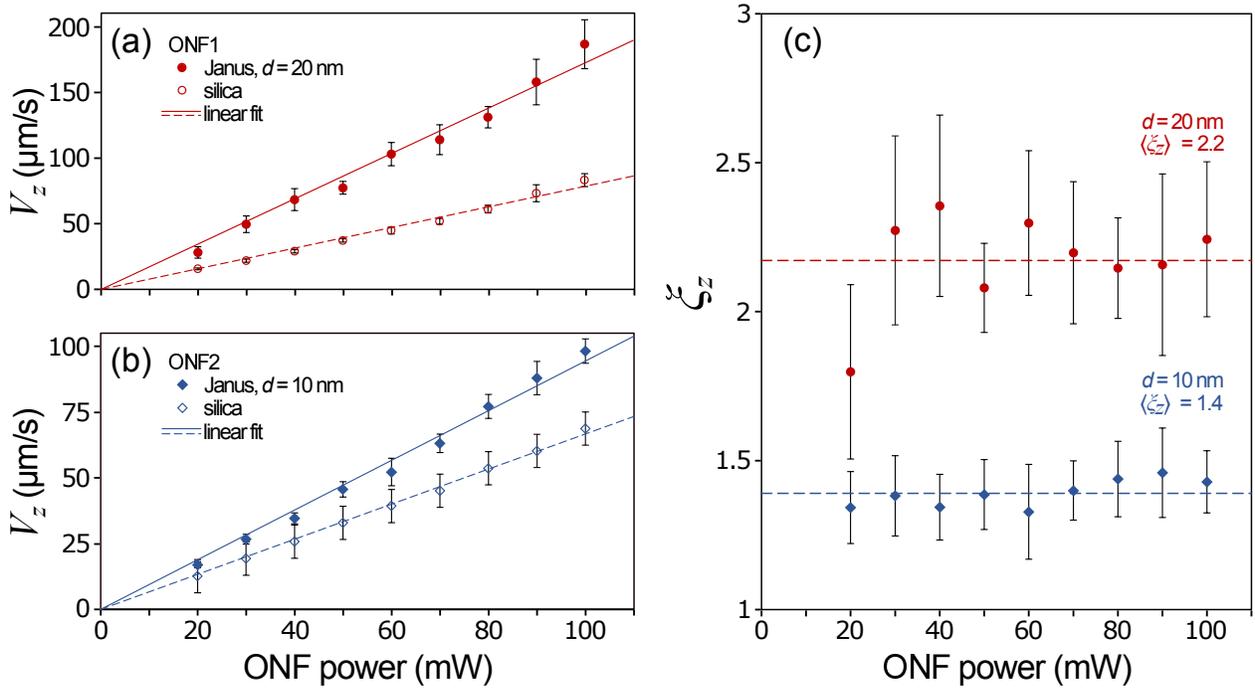}
\caption{(a,b)~Measured propulsion speed of $3~\mu$m silica and Janus particles near an ONF guiding a single mode polarized in the $yz$ plane ($\bf V$ input polarization) for a transmitted optical power ranging from 20~mW to 100~mW.  Two samples were used: ONF1 for $d=20$~nm and ONF2 for $d=10$~nm.   (c)~Propulsion enhancement factors, with the horizontal lines indicating the mean values, $\langle \xi_z \rangle$.}
\label{fig:figEnhance}
\end{figure}

The results of our quantitative study of the ONF-mediated propulsion of JPs is summarized in Fig.~\ref{fig:figEnhance}. As evident in panels~(a) and~(b), gold-coated particles consistently move faster than uncoated ones, for both values of the coating thickness, $d$, and for every power value in the explored range. Moreover, as confirmed by the well-fitting linear regression with zero intercept, the propulsion speed, $V_z$, is proportional to the transmitted optical power. When JPs are manipulated by an ONF, they can be viewed as standard dielectric microspheres, albeit with an enhanced effective polarizability. As a brief qualitative test, we increased the ONF transmitted power up to 200~mW and we still observed stable radial trapping of JPs and uniform propulsion following the same linear trend. In the experiments presented herein, we limited the  power  to 100~mW because performing accurate particle manipulations at higher powers (hence higher speeds) was too challenging for our experimental protocol.

To characterize the propulsion enhancement for the two kinds of JPs used, we calculated the mean enhancement factor, $\langle \xi_z \rangle=\langle V_{{\rm JP},z} \rangle / \langle V_{{\rm SP},z} \rangle$, for each power value. Here $\langle...\rangle$ stands for averaging over the recorded speed-position series (Figure~\ref{fig:figTraj}). Error bars in Figure~\ref{fig:figEnhance}c are calculated as $\pm\langle \xi_z \rangle \sqrt{[\sigma(V_{{\rm JP},z})/\langle V_{{\rm JP},z} \rangle ]^2+[\sigma(V_{{\rm SP},z})/\langle V_{{\rm SP},z} \rangle ]^2}$, where $\sigma(...)$ is the 50\% standard deviation range for the given speed-position series. The mean enhancement values over the whole range of ONF power are 2.2 and 1.4 for the 20- and 10-nm-thick gold coating, respectively.

\begin{figure}
\centering\includegraphics[width=\linewidth]{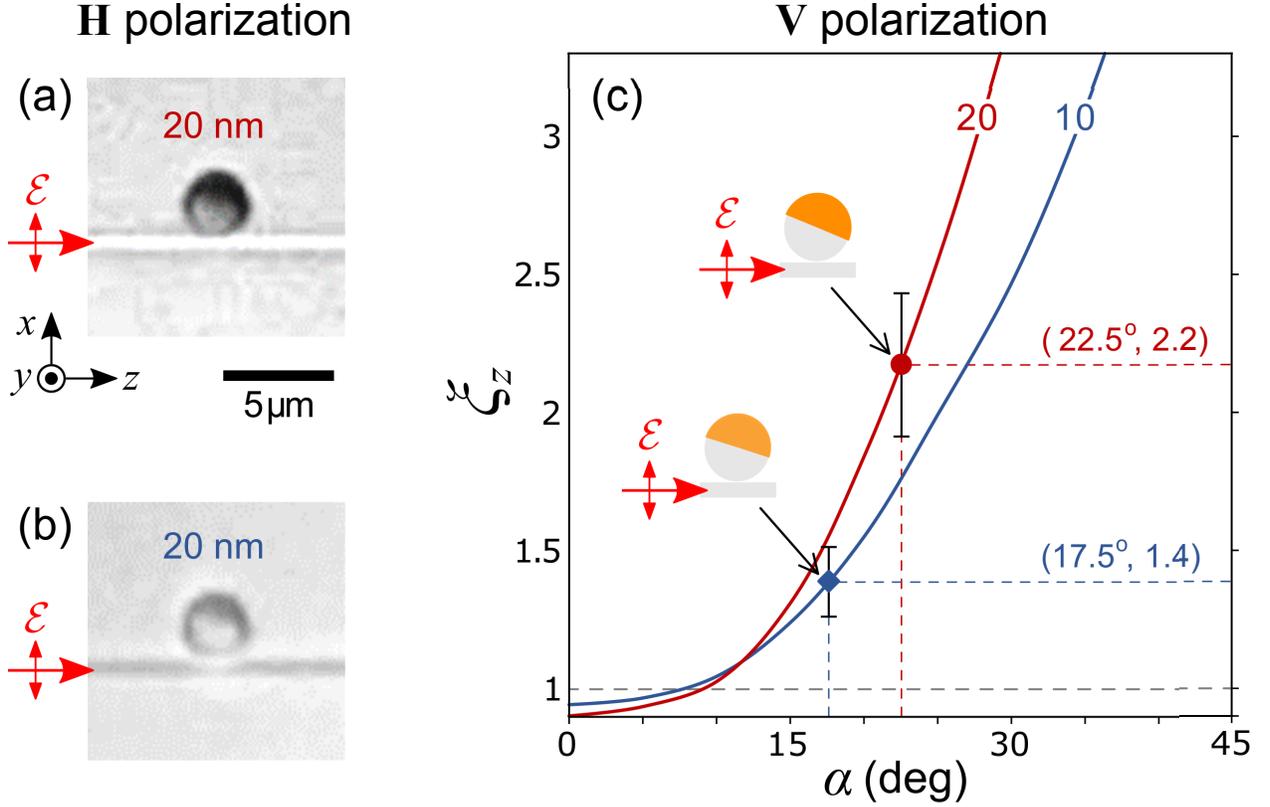}
\caption{(a,b)~Typical camera images of a JP with (a)~20- and (b)~10-nm-thick gold coating, in the process of being propelled at $\bf H$ input polarization and 100~mW of transmitted optical power (Supporting Information Video~S3). (c)~Simulated (curves) and measured (markers) propulsion enhancement factors versus the gold cap orientation angle, in the experimentally relevant range. Sketches represent the corresponding model configurations.}
\label{fig:figCap}
\end{figure}

\subsection{Gold cap orientation}
Now let us consider how the experimental results compare with the simulations (Figure~\ref{fig:figComsol}). Since we did not observe any signature of thermal effects near the ONF, we assume that the key factor defining the optical force enhancement is the orientation of the gold cap,~$\alpha$. To see the cap clearly, we had to set the input polarization to be horizontal, $\bf H$, in which case the particle was pulled toward the $xz$ plane where the mode has the highest intensity. Note that, to keep the particle in this plane, we had to use high optical power; otherwise the particle motion would sag along the $y$ axis under the transverse action of  gravity,~$\bf g$. For this reason, we have conducted the entire quantitative study of the gold cap orientation using $\bf V$ input polarization.

As evident from the typical images shown in Figure~\ref{fig:figCap}a,b, the cap tends to tilt in the direction of mode propagation, thus rendering $0<\alpha<90$. Assuming this tilt to be the same for $\bf H$ and $\bf V$ polarization states, we can link the measured propulsion enhancement, $\xi_z$, with $\alpha$, as shown in Figure~\ref{fig:figCap}c. Although the corresponding ONF+JP sketches visually agree with the microscope images (Figure~\ref{fig:figCap}a,b), we cannot claim that such images allow us to measure the tilt angle with a decent level of precision. Still, the observed forward tilt of the metallic cap is clear and reproducible. Given this effect, an ONF-propelled JP calls for an analogy with a vessel pushed by the pressure of wind on its sail. Our gold-coated (with ``sail'') silica particles tend to move very similarly to uncoated (no ``sail'') particles, except the ``sail'' gives a significant gain in the longitudinal speed.

\begin{figure}
\centering\includegraphics[width=1\linewidth]{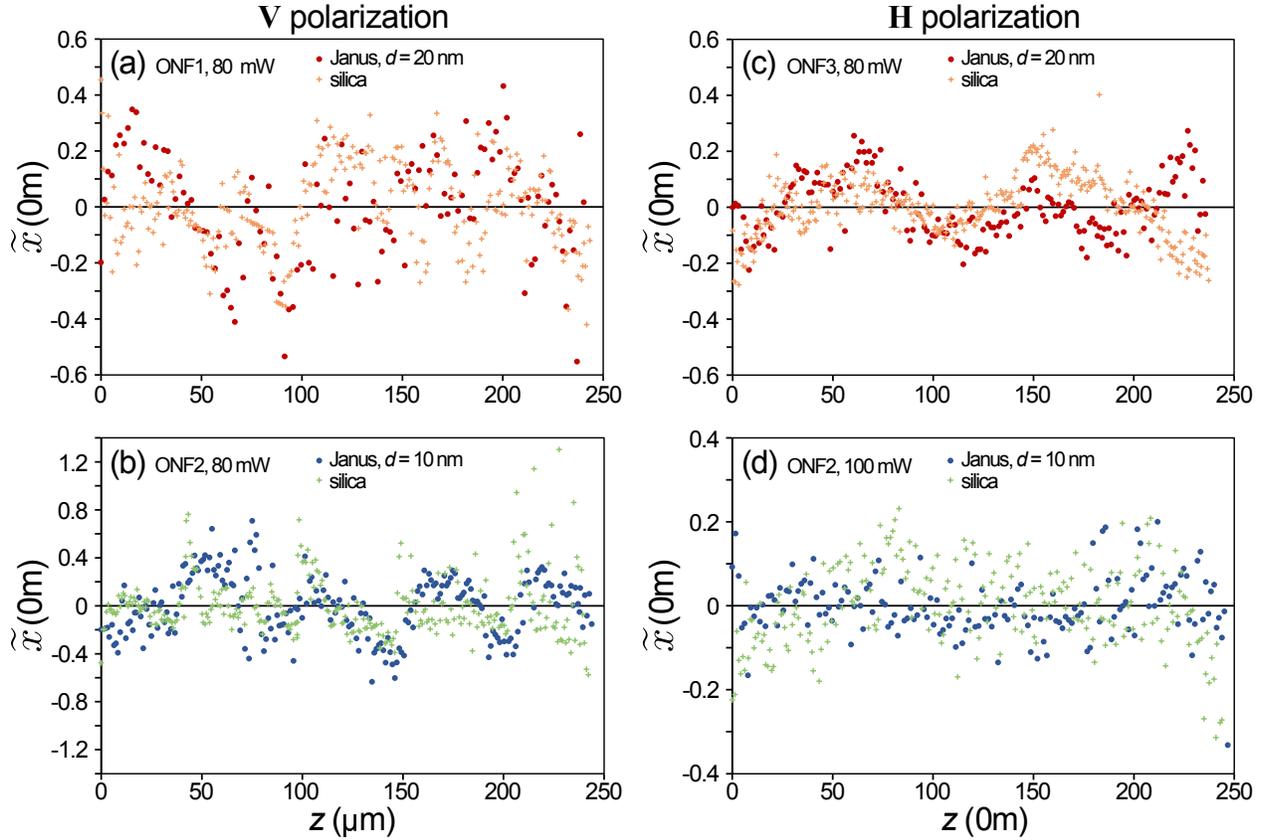}
\caption{Optical trapping dynamics of Janus and silica particles propelled by light near an ONF. (a,~b)~Typical time series for the reduced transverse coordinate of a JP compared to a SP in the same sample, coupled to a $\bf V$-polarized input beam~1. (c,~d)~Same for $\bf H$-polarized input. The optical power transmitted through the ONF is stated in the legends.}
\label{fig:figTrans}
\end{figure}

\subsection{Transverse localization}
Importantly,  a JP stays stably trapped in the transverse direction while being propelled along the ONF. In order to study this 2D optical trapping quantitatively, we compared the random motion of SPs and JPs along the $x$ axis. Typical position-time series are shown in Figure~\ref{fig:figTrans}, where $\tilde{x}(t)=x(t)-x_0(t)$ is the reduced transverse coordinate of the particle's center and $x_0(t)$ is the best linear fit for $x$ over the whole trajectory. We use $\tilde{x}$ instead of $x$ in order to compensate for possible tilts of the ONF with respect to the $z$~axis.

Following our assumption of thermal effects being negligible and assuming that the particle near an ONF is in a harmonic optical potential, we can roughly estimate the strength of the transverse trapping for JPs and SPs based on the equipartition theorem~\cite{neuman_rsi_2004}. It states that the trap stiffness (defined as the optical force per unit displacement) can be found from $\kappa_x = {\rm k_B}T/\sigma(\tilde{x})^2$, where ${\rm k_B}$ is Boltzmann’s constant, $T$ is the absolute temperature, and $\sigma(\tilde{x})^2$ is the positional variance. Table~1 summarizes our analyses of the data sets shown in Figure~\ref{fig:figTrans} by giving the measured propulsion enhancement factors, $\xi_z$, the standard deviation values for the transverse coordinate, $\sigma({\tilde x}_{\rm SP,\,JP})$, the corresponding transverse enhancement factors, $\xi_x = [\sigma({\tilde x}_{\rm SP})/\sigma({\tilde x}_{\rm JP})]^2$, and the simulated radial enhancement factors, $\xi_r$ (equivalent to $\xi_y$ in Figure~\ref{fig:figComsol}c).

\begin{table}
\centering
\caption{Analysis of the data series shown in Figure~\ref{fig:figTrans}}
\begin{tabular}{lrrrrrr}
            \hline
            {Polarization}&&\multicolumn{2}{c}{$\bf V$}&&\multicolumn{2}{c}{$\bf H$}\\
            \cline{1-1}\cline{3-4}\cline{6-7}
            {Sample}&&{ONF1}&{ONF2}&&{ONF3}&{ONF2}\\
            {ONF power (mW)}&&{80}&{80}&&{80}&{100}\\
            {$d$ (nm)}&&{20}&{10}&&{20}&{10}\\
            {$\xi_z$, experiment}&&{2.17}&{1.41}&&{1.54}&{1.47}\\
            {$\sigma({\tilde x}_{\rm SP})$ (nm)}&&{152}&{307}&&{175}&{97}\\
            {$\sigma({\tilde x}_{\rm JP})$ (nm)}&&{193}&{250}&&{105}&{70}\\
            {$\xi_x$, experiment}&&{0.62}&{1.51}&&{2.78}&{1.92}\\
            {$\xi_r$, simulation}&&{0.85}&{0.91}&&{0.86}&{0.91}\\
            \hline
        \end{tabular}
\label{tab1}
\end{table}

As one can see from Table~\ref{tab1}, $\xi_x>1$ in all cases except for JPs with $d=20$~nm propelled under $\bf V$ polarization. Strictly speaking, the experimental measurements can only be compared to simulations for $\bf H$ polarization, in which case the $x$ direction is truly radial for the 2D-trapped particle (gravity-induced displacement is neglected for high power). For both values of the gold thickness, $\xi_x>1$, which means that the metallic coating leads to the particle position being more localized by light.  This is an encouraging result for prospective waveguide-mediated optomechanical tools for JPs. However, this does not agree with the model prediction in the observed range of $\alpha$ (Figure~\ref{fig:figComsol}e). We attribute this discrepancy to thermal forces, which are not considered in our limited model. Such forces are known to push silica-gold JPs in water away from the higher-temperature region created around the cap~\cite{jiang_prl_2010,mousavi_softMat_2019}, thus JPs in our experiments are pushed to the fiber surface stronger than one would expect from optical forces alone. In order to improve the accuracy of the theoretical analysis, one could consider accounting for thermophoresis, heat-induced spatial variations of viscosity, the electrical potential map at the slipping plane~\cite{rashidi_jcp_2017}, and likely nonuniformities of the gold cap~\cite{zong_acsNano_2015,rajupet_pre_2021}. All these aspects are beyond the scope of this study. 

\section{Conclusion and outlook}
We have reported on the optical manipulation of metallo-dielectric Janus particles (silica microspheres half-coated with a thin layer of gold) in the evanescent field of an optical nanofiber. Our experiments demonstrate that such a particle is stably trapped at the nanofiber surface and steadily propelled along the mode propagation direction, with speeds exceeding those typically measured for uncoated silica beads of the same size. Specifically, we registered  speed enhancement factors of 1.4 and 2.2 for gold coating thicknesses of 10~and 20~nm, respectively. These factors were independent of the optical power, hence Janus particles in the evanescent field exhibit largely the same light-induced dynamics as standard dielectric beads: (i)~steady propulsion with a speed proportional to the optical power and (ii)~negligibly small thermal effects (at least as far as the propulsion is concerned). For a 20-nm-thick coating, the average speed of a 3-$\mu$m particle was approximately 2~$\mu$m/s per 1~mW, an equivalent of 133~body length/s for the highest power of 200~mW in our tests. This is over 5~times faster than the optical propulsion of similar Janus particles in free space~\cite{dong_acsNano_2016}. Comparing our experimental and theoretical results, we link the propulsion enhancement to the orientation of the gold cap, which is consistently recorded to be facing away from the nanofiber and toward the direction of motion, effectively acting as an optical ``sail’’. Our model predicts that the enhancement could be up to an order of magnitude higher if the cap were to tilt closer to the evanescent field. In this regard, it would be interesting to try geometries where the metal covers more than half of the particle, such as for a ``nanocoral’’~\cite{wu_small_2010}.

Aside from the efficient propulsion, we have found that Janus particles exhibit strong optical trapping in the transverse direction, with the stiffness typically exceeding that of uncoated silica beads. This feature enhances the precision of the optical manipulation and is therefore highly beneficial for applications of Janus particles as micro- or nanoscale actuators or transporters. We envisage further development of waveguide-based technologies with Janus particles, in particular contributing to the prospective ``rail’’-style microfluidic designs~\cite{frenzel_labChip_2017, loo_jom_2021}.



\section{Methods}
\subsection{Numerical simulations}
To calculate the optical force acting on the Janus particle, we first simulate the distributions of surrounding electric and magnetic fields by solving the scattering problem via the finite element method (COMSOL 5.5 Multiphysics package). These fields are then used to define the time-independent Maxwell stress tensor, numerical integration of which over the area surrounding the particle gives the total optical force. Our model consists of a two-dimensional $y$-$z$ view of a particle located at 10~nm distance from the surface of a 0.7-$\mu$m-thick silica fiber, centered in the simulation domain of $z\times y = 21\,\mu{\rm m}\times11\,\mu{\rm m}$, padded along $y$ by 1-$\mu$m-thick perfect matching layers to simulate open boundaries. An in-plane polarized fundamental mode with the wavelength of 1.064~$\mu$m  propagates through the fiber towards $z>0$. For all calculations, the total propagating power of the guided light was normalized to 1~mW. JPs are modeled by a core 3-$\mu$m silica circle with concentric semi-circular caps of titanium and gold, as shown in Figure~\ref{fig:figComsol}a. The radial force, ${\bf F}_y$, originates from the transient polarisation of the trapped particle in a non-uniform field. The longitudinal force, ${\bf F}_z$, is responsible for propelling the particles along the fiber. We set the outline for integration of the Maxwell stress tensor to be positioned at 5~nm from the outermost edge of the particle. The refractive indices of silica structures and the surrounding water were taken as 1.45 and 1.33, respectively. Optical properties of metals were defined using the Brendel-Bormann model~\cite{rakic_ao_1998}.

\subsection{Particle preparation}
Silica-gold JPs were prepared following the evaporation coating method~\cite{nedev_acsPhot_2015}. First, a loosely packed monolayer of silica microspheres ($2R_{\rm p}=3.13\pm0.2\mu$m by Bangs Laboratories, Inc.) was formed on a glass substrate by drop-casting them in an ethanol suspension and drying under ambient conditions. Second, the substrate was loaded into an electron beam evaporator (MEB550S2-HV by PLASSYS-BESTEK) and sequentially coated with a 5-nm-thick adhesion layer of titanium and a 10- or 20-nm-thick layer of gold. Finally, JPs were detached from the substrate by sonication for 15~minutes in Milli-Q water. This stock suspension of particles was further diluted by Milli-Q water to achieve a convenient concentration (about 10 particles in the full field-of-view, FoV) for the optomechanical experiments.

\subsection{Optical nanofiber sample fabrication}
The ONFs were fabricated via controlled heating and pulling~\cite{ward_rsi_2014} of standard step-index single-mode optical fibers (SM980G80 by Thorlabs, Inc.). The tapered regions had a linear profile with a half apex angle of 3~mrad. Adiabaticity of the tapering was verified by {\it in situ} monitoring of the transmission of a probe laser beam ($1.064\,\mu$m wavelength) coupled to one of the fiber pigtails; only the samples having $>98\%$ transmission were used in the optomechanical experiments. The ONF waist region was engineered to be a 2-mm-long cylinder with a radius of $0.7\,\mu$m (measured to be $0.66\pm0.08\,\mu$m in practice~\cite{tkachenko_optica_2020}). A freshly pulled ONF was suspended over a 0.15-$\mu$m-thick glass cover slip on pieces of 0.08-$\mu$m-thick tape and secured by UV adhesive glue (NOA81 by Norland Products, Inc.). After depositing about 0.3~mL of particle suspension in Milli-Q water onto the ONF waist and taper regions, we covered them with a second glass cover slip, supported on 1-mm-thick polymer spacers. The sample had its sides open to the atmosphere, thus all the measurements on each were performed in one session, in order to avoid detrimental effects of water evaporation.

\subsection{Optical setup}
Samples were mounted on a three-axis mechanical stage (MAX313D by Thorlabs, Inc.). Beams~1-3 in Figure~\ref{fig:figSetup}a were collimated, non-interfering, linearly polarized Gaussian beams from the same laser source (Ventus by Laser Quantum Ltd.). Beams~1 and~2 were coupled to the fiber pigtails of the sample. We used the guided modes polarized in the vertical ($yz$) or horizontal ($xz$) planes. To this end, we set the input polarization state in each of the beams 1 and 2 to vertical ($\bf V$, along $y$) or horizontal ($\bf H$, along $x$) by means of half-wave plates (not shown) and performed a partial polarization compensation procedure by mapping ${\bf H}\to{\bf H}$ at the ONF waist~\cite{lei_prappl_2019}. The procedure relies on the independent adjustment of two quarter-wave plates in each compensator, PC1 and PC2, while imaging the laser light scattered from the waist~\cite{tkachenko_jo_2019}.

The optical tweezers for trapping and delivering individual particles to an ONF was realized by tightly focusing  beam~3 via a water-immersion objective lens (Zeiss Plan-Apochromat, $63\times/1.00$w). Transmission imaging of the sample under K{\"o}hler illumination was performed through the same objective, which projected the image (Figure~\ref{fig:figSetup}c) onto a video camera (DCC3240C by Thorlabs, Inc.) through a short-pass dichroic mirror and a tube lens with 75-mm-long focal distance. To ensure correct timing of the camera operated by the native Thorlabs software, we cropped the FoV to about 1/5 of its full size (fitting the fiber and the particle) and set the pixel clock to be under 20~MHz, with a frame rate of about 70~frames per second.

The optical power in the tweezers was measured in real time by a calibrated photodetector (PDA10A2 by Thorlabs, Inc.) capturing a small portion of beam~3 passing through the dichroic mirror. To estimate the optical power at the ONF waist, we deflected a few percent of the transmitted light from the driving beam~1 by a glass wedge to another calibrated photodetector (also PDA10A2 by Thorlabs, Inc.). Although we cannot measure the power at the waist directly, this transmission-based measurement is expected to give an accurate estimate, because every sample ONF had over 98\% transmittance as-fabricated, and we expect most of the losses to occur due to radiation modes launched from the down-taper region.
The power was measured just before bringing a particle into contact with the ONF and checked after releasing the particle at the end of a propulsion cycle. We found the power to be stable within $\pm2$\%. Real-time measurements of the ONF power were inaccurate, because the transmitted power was seen to vary in about $\pm20$\% range due to scattering by the particle. In our experiments, the transmitted ONF power varies from 20~mW to 100~mW with the step of~10~mW.
While working with each sample, we checked the efficiency of fiber coupling for beam~1 (about 65\% for an average sample) at least once in an hour, as a control for the ONF cleanliness and the polarization accuracy.

\subsection{Measurement procedure}
Before starting the propulsion experiments, we first determined the thinnest and the most uniform part of the ONF waist by optomechanical probing of its evanescent field. For this purpose, we selected a SP with the optical tweezers, brought it to the ONF, unblocked   beam~1 so the particle would be trapped near the ONF, and switched off the tweezers by blocking beam~3. Then we observed the particle's motion and moved the sample along $z$, until we found the position where $V_z$ was the highest and the most uniform across the FoV.

A typical cycle of the propulsion measurements consisted of the following steps: (i)~unblock beam~3 to form a 3D optical trap around the centre of the FoV, (ii)~trap a JP at the bottom glass slide, bring it up to the ONF, (iii)~unblock  beams~1,2 and block beam~3 to transfer the particle into the evanescent field trap, (iv)~block  beam~1 to propel the particle toward $-z$; (v)~once it has reached the left edge of the FoV, unblock  beam~1, block  beam~2, and record the particle's propulsion all the way to the right edge of the FoV, (vi)~block all beams and repeat steps (i)-(v) for a SP. In practice, compared to SPs, JPs were harder to find and to move laterally, thus we tried to recycle JPs by propelling them back to the centre of the FoV with  beam~2 (Supporting Information Video~S1). Still, each data set required 2-4 different JPs which allowed us to assess the reproducibility of the results within each sample. The particle's speed was obtained by analyzing video recordings frame by frame, where the particle's position was defined as the centre of the circle fitted to its outline.

\begin{acknowledgement}
The authors thank K. Karlsson, M. Ozer, and the Engineering Support Section of Okinawa Institute of Science and Technology Graduate University (OIST) for technical assistance. This work was funded by OIST and the Japan Society for the Promotion of Science (JSPS) KAKENHI (Grant-in-Aid for JSPS Fellows) 18F18367. G. T. was supported by a JSPS International Research Fellowship (Standard) P18367.

\end{acknowledgement}

\begin{suppinfo}
Demonstration videos; S1 shows the typical behavior of a JP in optical tweezers and in the evanescent field of an ONF; S2 presents the source video recordings for the longitudinal position-time series plotted in Figure~\ref{fig:figTraj}a; S3 shows the test recordings of JPs and SPs propelled at horizontal polarization, where the gold cap orientation is visible.

\end{suppinfo}

\bibliography{ref}

\end{document}